\documentclass{llncs}
\usepackage{booktabs} 
\usepackage{comment}
\usepackage{graphicx}
\usepackage[lofdepth,lotdepth]{subfig}
\usepackage{cellspace}
\usepackage{wrapfig,picins}
\usepackage[export]{adjustbox}

\usepackage[sorting=none, maxnames=50]{biblatex}
\usepackage{amsmath}
\usepackage{color}
\usepackage{url}
\usepackage{booktabs} 
\usepackage{amsfonts}
\DeclareGraphicsExtensions{.pdf,.png,.jpg}
\usepackage[english]{babel}
\usepackage[table,xcdraw]{xcolor}
\usepackage[section]{placeins}

\usepackage{listings}
\usepackage{xcolor}
\definecolor{codegreen}{rgb}{0,0.6,0}
\definecolor{codegray}{rgb}{0.5,0.5,0.5}
\definecolor{codepurple}{rgb}{0.58,0,0.82}
\definecolor{backcolour}{rgb}{0.95,0.95,0.92}
\lstdefinestyle{mystyle}{
    backgroundcolor=\color{backcolour},   
    commentstyle=\color{codegreen},
    keywordstyle=\color{magenta},
    numberstyle=\tiny\color{codegray},
    stringstyle=\color{codepurple},
    basicstyle=\ttfamily\footnotesize,
    breakatwhitespace=false,         
    breaklines=true,                 
    captionpos=b,                    
    keepspaces=true,                 
    numbers=left,                    
    numbersep=5pt,                  
    showspaces=false,                
    showstringspaces=false,
    showtabs=false,                  
    tabsize=2
}
\begin{document}

\frontmatter          
\pagestyle{empty}  

\title{IoT to monitor people flow in areas of public interest}
%
%
 \author{
 Damiano Perri\inst{1,2} $^{ORCID: 0000-0001-6815-6659}$\newline 
 Marco Simonetti\inst{1,2} $^{ORCID: 0000-0003-2923-5519}$\newline 
 Alex Bordini\inst{2} $^{ORCID: 0000-0002-4445-7876}$ \newline 
 Osvaldo Gervasi\inst{2} $^{ORCID: 0000-0003-4327-520X}$ 
 }
 
\institute{
University of Florence, Dept. of Mathematics and Computer Science, Florence, Italy \and University of Perugia, Dept. of Mathematics and Computer Science, Perugia, Italy
}
\titlerunning{IoT to monitor people flow in areas of public interest} 
\authorrunning{D. Perri, M.  Simonetti, A. Bordini, and O. Gervasi} 

\maketitle

\begin{abstract}
The unexpected historical period we are living has abruptly pushed us to loosen any sort of interaction between individuals, gradually forcing us to deal with new ways to allow compliance with safety distances; indeed the present situation has demonstrated more than ever how critical it is to be able to properly organize our travel plans, put people in safe conditions, and avoid harmful circumstances. 
The aim of this research is to set up a system to monitor the flow of people inside public places and facilities of interest (museums, theatres, cinemas, etc.) without collecting personal or sensitive data.
Weak monitoring of people flows (i.e. monitoring without personal identification of the monitored subjects) through Internet of Things tools might be a viable solution to minimize lineups and overcrowding. 
Our study, which began as an experiment in the Umbria region of Italy, aims to be one of several answers to automated planning of people's flows in order to make our land more liveable.
We intend to show that the Internet of Things gives almost unlimited tools and possibilities, from developing a basic information process to implementing a true portal which enables business people to connect with interested consumers. 

\end{abstract}


\section{Introduction}
One of the characteristics related to the pandemic of COVID19 as a result of social distancing is that it has forced institutions to reduce the maximum acceptable number of simultaneous visitors.
This situation generated difficulties from the organisational point of view of both the structures and the tourists who wanted to visit certain areas of cultural and tourist interest.
The words "Internet of Things" (IoT) refer to a collection of objects that gain intelligence through the internet, exchanging data about themselves and gaining access to information from other sources \cite{ATZORI20102787, mainetti2011evolution}.


The goal of this research is to maximize the potential of IoT devices to collect particular data by combining privacy with useful information, simplicity of use with usefulness, economy with power, and versatility with accuracy.
This project will aim at leveraging existing infrastructure to develop the concept of indoor tracking, i.e. the ability to provide information on the number of people presently inside a building, public or private, and selectively providing the details publicly available.

The system we are proposing is made up of three major components: acquisition devices, which deal with understanding how the flow varies over time based on the impulses they receive, coordination devices, which deal with managing all of the information received, and presentation devices, which make the previously acquired information available to the public.

\section{Related works}
The effort to automate work and manufacturing processes that began in the nineteenth century has never truly ended, due to the assistance of increasingly efficient and performing technology. 
Today's strategy is to make machines increasingly autonomous and intelligent\cite{DBLP:conf/iccsa/BenedettiPSGRF20,DBLP:conf/iccsa/PerriSLLG20,DBLP:conf/iccsa/PerriLGTV19}, rather than highly reliant on human supervision\cite{mcfarlane2003auto, zhong2017intelligent}.
This has required interaction with a complex world, which necessitates the development of numerous external inputs on which to take decisions\cite{DBLP:conf/iccsa/FranzoniTPP19,DBLP:conf/iccsa/LaganaGTPF18}.
As a result, the Internet of Things has become a significant paradigm in a variety of fields, including manufacturing\cite{yao2017intelligent, wang2021industrial}, remote control\cite{adhya2016iot, pallavi2017remote} and monitoring\cite{ghosh2016remote, na2016iot}, home automation\cite{kodali2016iot, pavithra2015iot}, gaming\cite{santucci2020immersive} and Virtual and Augmented Reality\cite{hu2021virtual,DBLP:conf/iccsa/SimonettiPAG20}.
Our concept is based on the basic requirement to track the flow of individuals moving through restricted settings for a variety of reasons, including real-time monitoring and control to minimize congestion, and statistical assessment for economic and logistical considerations.
Although the goal is the same, the methodologies used over time have been the most varied: RFID techniques\cite{DBLP:conf/iccsa/GervasiFMPS19}, ultrasound sensors\cite{therib2020smart}, PIR sensors\cite{sruthi2019iot}, closed-circuit TV (CCTV)\cite{saon2020cloud}, face recognition by Artificial Intelligence \cite{nag2018iot,DBLP:conf/iccsa/BiondiFGP19, othman2018face}, post-processed data local memories or in Cloud Services\cite{cai2016iot}.

The use of IoT has always proven to be a simple answer, given the low cost of the components, which certainly favors the implementation of a do-it-yourself solution.
The method we presented is intended to be simple, both in terms of design and implementation, as well as in terms of simplicity of use and with a low learning curve.

\section{The Architecture of the system}
Our system is tasked with monitoring and tracing the movement of people passing through a \textit{location}, which is a confined environment such as a public or private structure.
This data is very useful in ensuring the safety and security of people in the area, and it is controlled differently by those who have the information, and decides what to publish, and those who utilize it, i.e. the end user.

\begin{figure}[h!]
    \centering
    \includegraphics[width=0.5\linewidth]{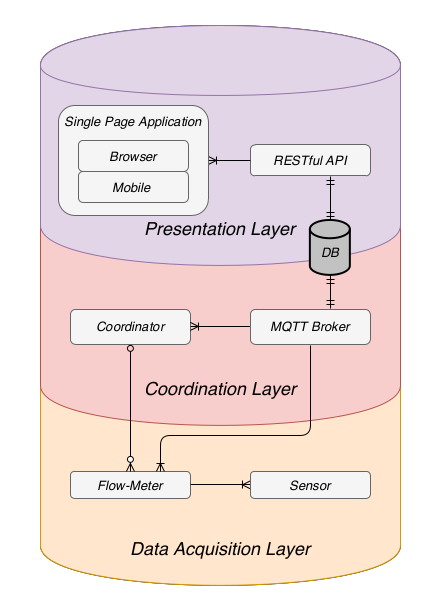}
    \caption{System Architecture scheme}
    \label{fig:fig01}
\end{figure}

To ensure the generic collection of data, useful for guaranteeing service, and to realize the anonymity of individual people, a scalable stack architecture, as shown in Figure \ref{fig:fig01}, is divided into three logical levels: \textit{data acquisition layer}, \textit{data coordination layer} which has the function to improve a correct exchange of information across devices, and \textit{presentation layer} needed to allow the flexible use of information in manageable formats.

\subsection{Data Acquisition Layer}
This layer contains every object related to the collection and conversion of sensor data into processable information from higher layers.

The first critical component of our research is a collection of flow acquisition devices known as \textit{Flow-Meters}.
They can be heterogeneous or homogeneous: in some situations, using different sensors may be convenient to get a wide overview; 
in others, using sensors of the same kind may enhance overall accuracy or area detection.
The first scenario entails the employment of numerous sensors of various sorts to improve detection possibilities, utilizing the plurality of signals received to manage the complexity of the surrounding environment.
In the second case, the detection system is implemented using a single type of sensor, like an emitter and a receiver photo-diode; such a simple structure was initially considered but later rejected because it is incapable of effectively describing contingent situations, as the direction of the person.

\begin{figure}[h!]
    \centering
    \subfloat[Thermal Camera Sensor]{
        \includegraphics[width=0.261\linewidth]{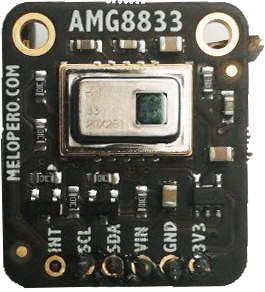}
        \label{fig:subfig1}}
    \subfloat[Grid-EYE matrix sensor]{
        \includegraphics[width=0.35\linewidth]{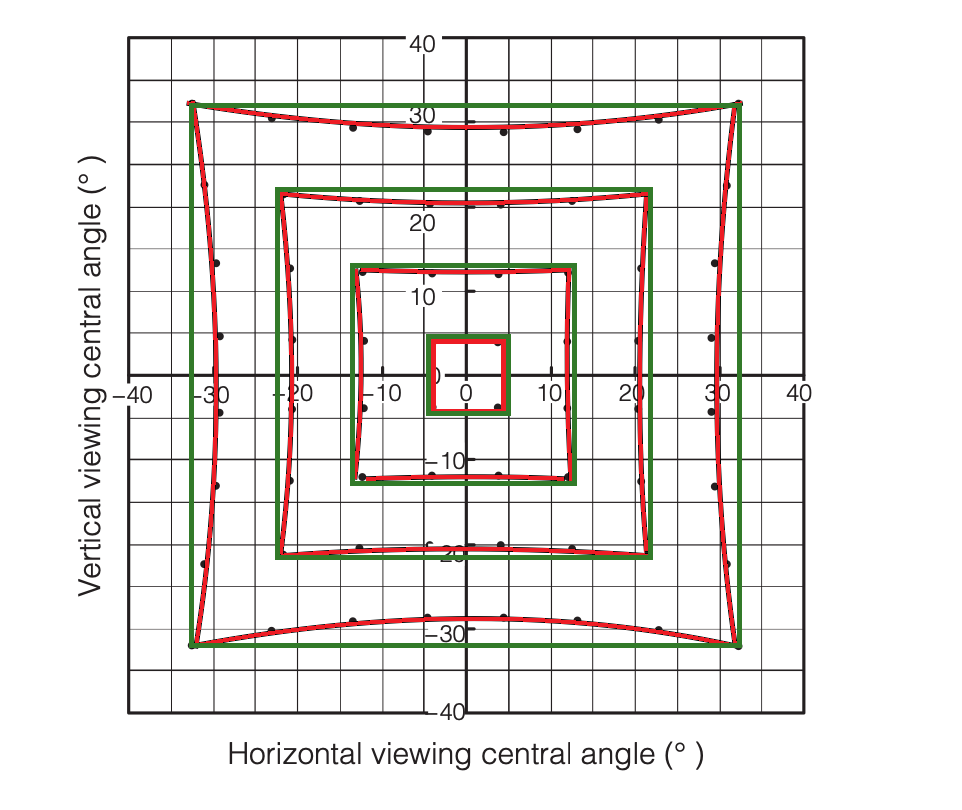}
        \label{fig:subfig2}}
    \subfloat[Thermoscanner]{
        \includegraphics[width=0.35\linewidth]{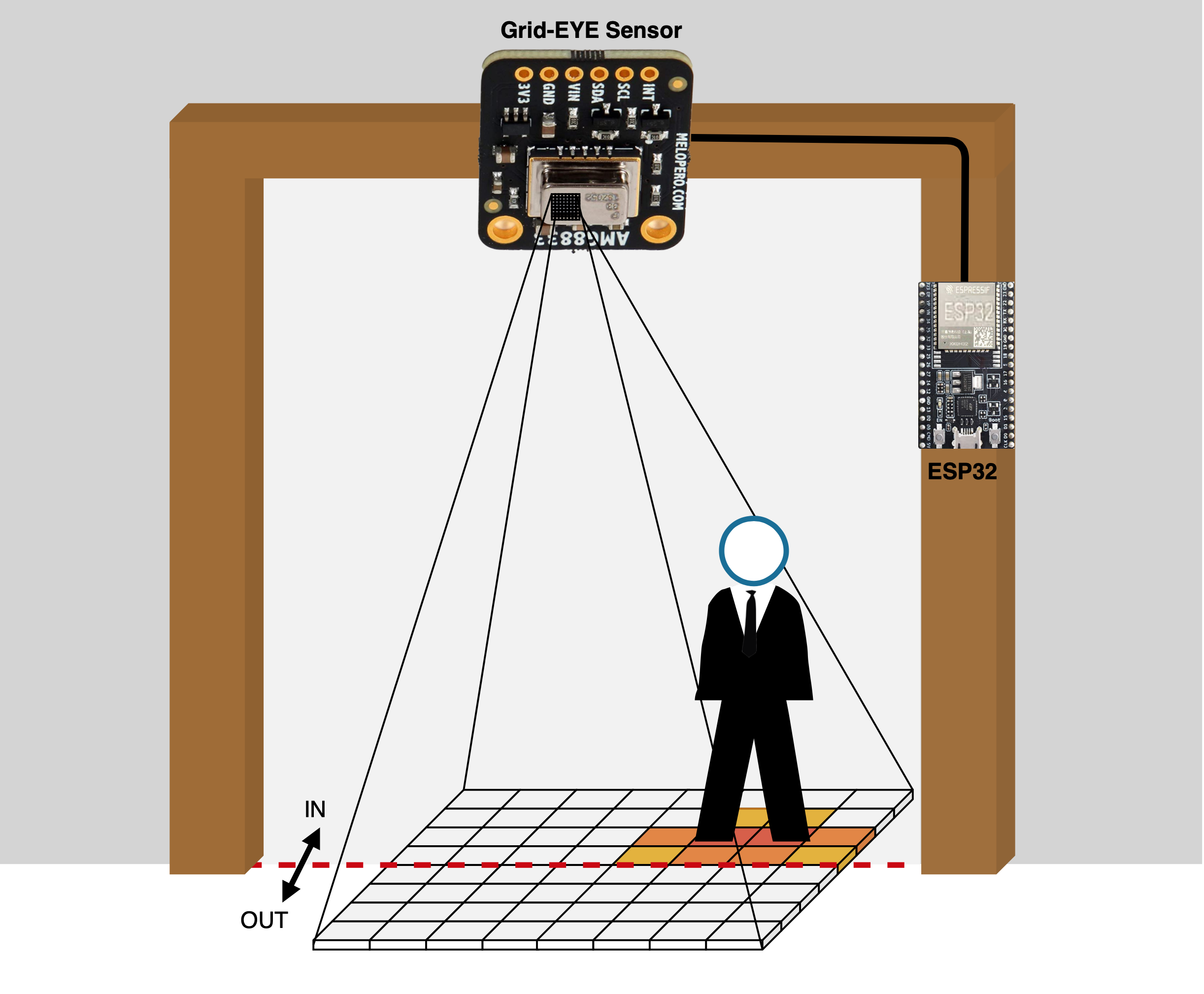}
        \label{fig:subfig3}}
    \caption{GRID-EYE matrix sensor}
    \label{figure:grideye}
\end{figure}

To prevent issues with correctly recognizing the passing of one or more persons, we ultimately choose a thermal camera sensor, best known as \textit{Grid-EYE matrix sensor}, shown in Figure \ref{fig:subfig1}.
This is one of the most common option in the field of Indoor Localization.
It is a tiny detector, less than a centimeter in size, that creates a two-dimensional 8x8 matrix, able to detect the average temperature for the section of the plane that is projected to the ground using lasers similar to those used in thermoscanners - Figure \ref{fig:subfig3}.

The size of the detectable plane sections is obviously determined by the distance between the device and the ground: the higher the device's positioning height, the larger the effect on the single viewing angle measurements, i.e. the incidence of each laser sensor on the ground.
The area covered by sensor and relative geometrical distortion is shown in Figure \ref{fig:subfig2}.

To manage the incoming flow from the different sensors, we utilize an ESP32 micro-controller, which is responsible for processing the raw data and creating a version suitable for control devices.
It continually interrogates the sensor and extracts all of the information about the flow. 
The huge amount of RAM memory required for the management of our matrices needed a sophisticated micro-controller.

The sensor's output signal is a low-resolution thermal image.
It is accompanied with a lot of noise, which has been reduced by using interpolation techniques and filtering matrices.
Initially, we sought a method that attempted to make these passages less apparent, incorporating filters to tackle the problem; these filters are known as \textit{convolutive} in the context of digital image processing.
A convolutive filter is a kernel matrix that is applied to another matrix, in this case the one from the sensor, to produce an effect based on the values it assumes.
The initial working hypothesis was to blur the image; however, after implementing the algorithm, it was discovered that the benefits were minimal because the system was limited by the few pixels available.
As a result, the originally expected homogeneity would not have been achieved, and some information would have been lost, eliminated by the convolutive filter.
This indicated to us that the primary requirement was not to mix the pixels we already had, but to dilute them over a broader region.
Interpolation is a method that involves adding new pixels to existing ones and calculating their value using a reconstruction algorithm.
The technique we have chosen is one of the most often used tools for upscaling in photo editing, \textit{bicubic interpolation}, whereas the technique to filter pixels has been \textit{Background Subtraction} \cite{Kallur2014HumanLA}.

\begin{figure}[h!]
    \centering
    \subfloat[]{
        \includegraphics[width=0.5\linewidth]{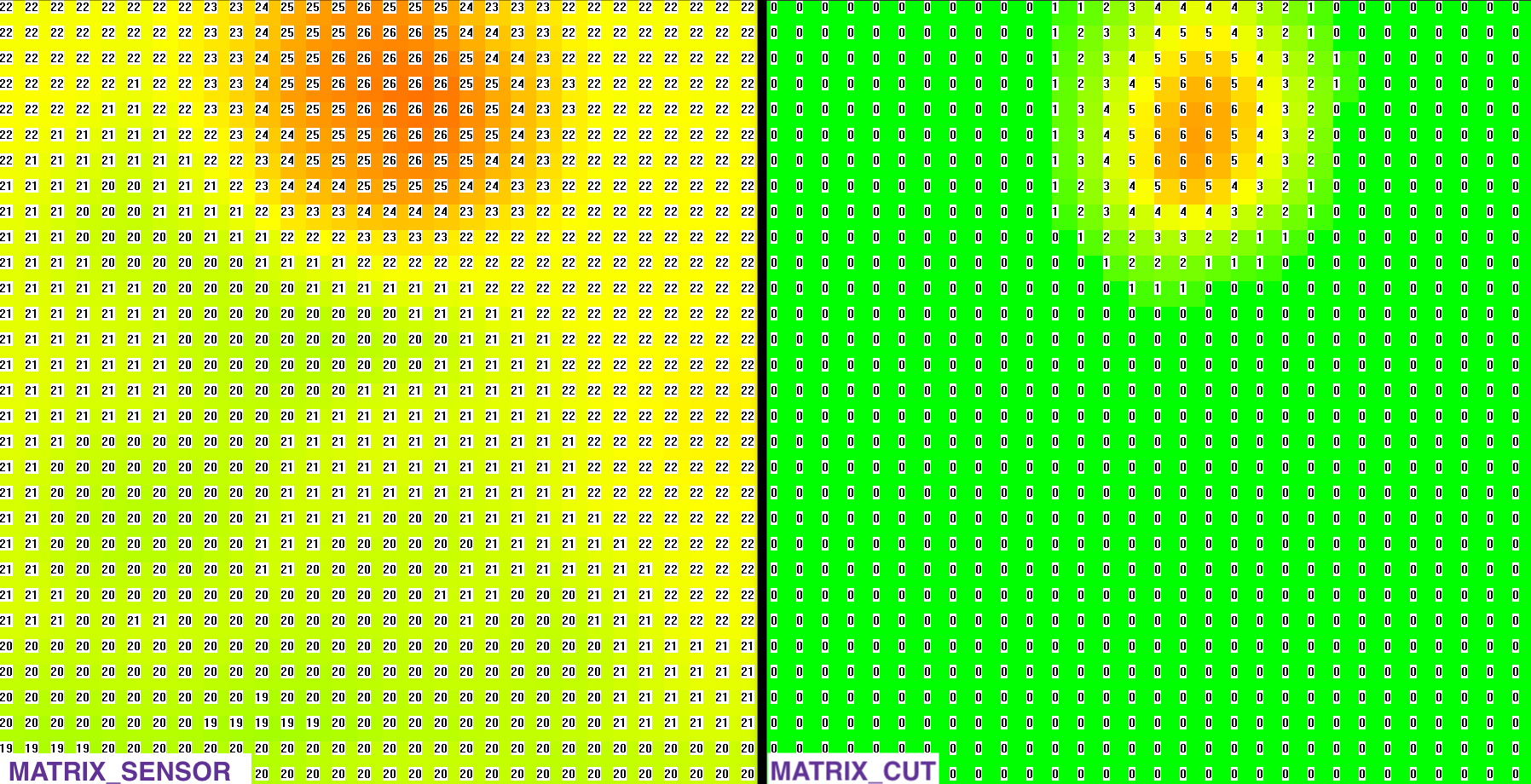}
        \label{fig:subfig4}}
    \hfill
    \subfloat[]{
        \includegraphics[width=0.3\linewidth]{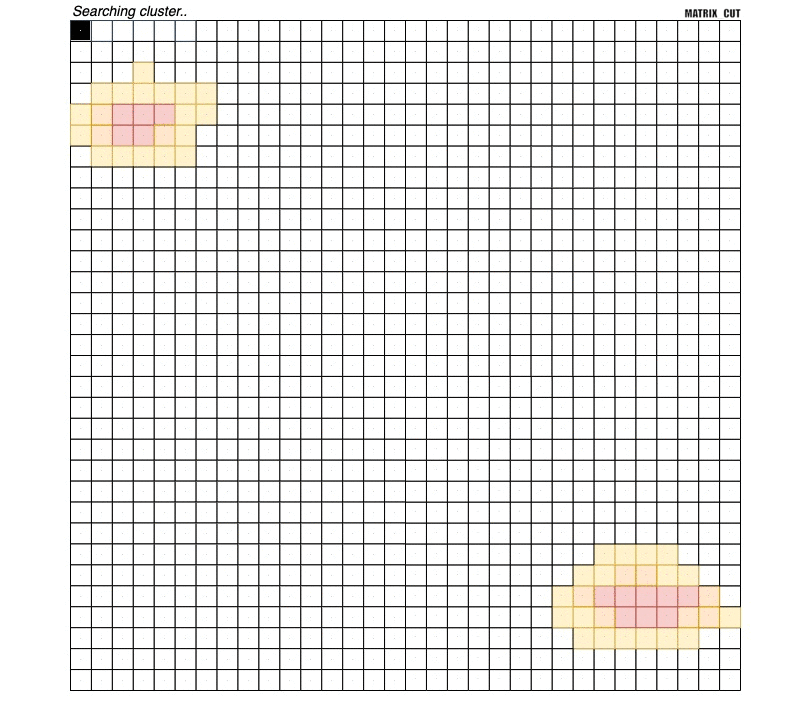}
        \label{fig:subfig5}}
    \caption{Image Processing}
    \label{figure:imageprocessing}
\end{figure}

Only when the noise has been adequately attenuated, the instrument's sensitivity can be modified using a threshold value.
This allows you to maintain the highest values of the matrix unchanged while attenuating the minimums and clearly identifying the individual clusters as shown in Figure \ref{fig:subfig4}.
We called \textit{cluster} a submatrix with variable shape and size, that numerically represents the area which the person can be projected on.

The matrix is then scanned to determine the size of the clusters.
The objective is to minimize these regions to a single point.
This point is their center of gravity, and acquire one for each individual in order to work on data distribution uniformity.
To do this, a Flood-Fill algorithm is used to extrapolate the individual clusters, and then their centers of gravity are computed using a weighted average ah shown in Figure \ref{fig:subfig5}.

\begin{figure}[h!]
    \centering
    \includegraphics[width=0.8\linewidth]{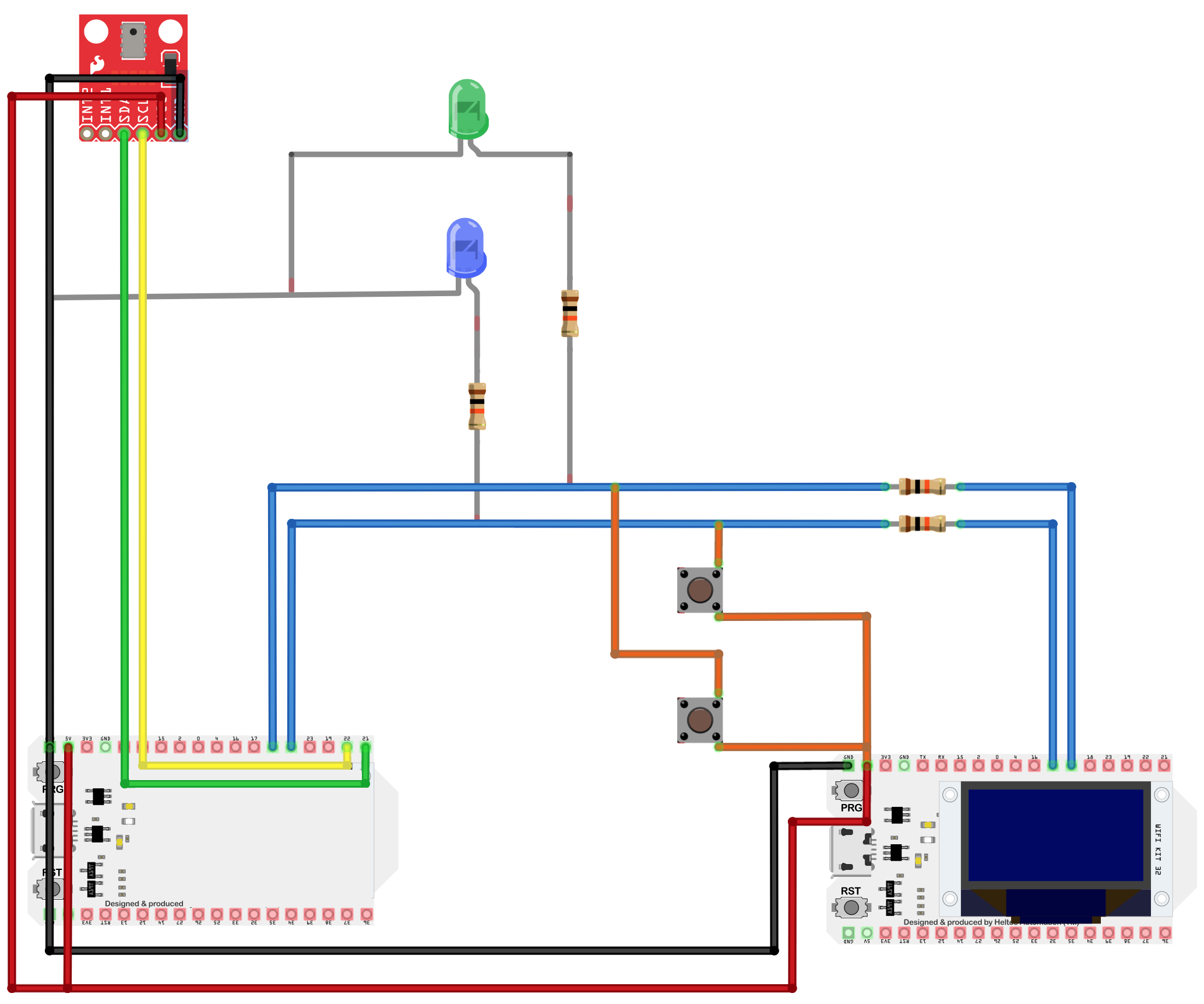}
    \caption{Wiring diagram for the circuit}
    \label{fig:elscheme}
\end{figure}

To convey the results of this level's computations to higher levels, the device requires a communication logic, capable of being simple and reliable.
Physical communication in binary form is used to support this, with increments and decrements communicated to the relevant pins through the electrical potential difference.
In this way a proper electrical diagram can help us to quickly provide feedback on functioning and faults as shown in Figure \ref{fig:elscheme}.

\subsection{Coordination Layer}
\begin{figure}[h!]
    \centering
    \subfloat[Interface among the MQTT broker and microcontrollers]{
        \includegraphics[width=0.9\linewidth]{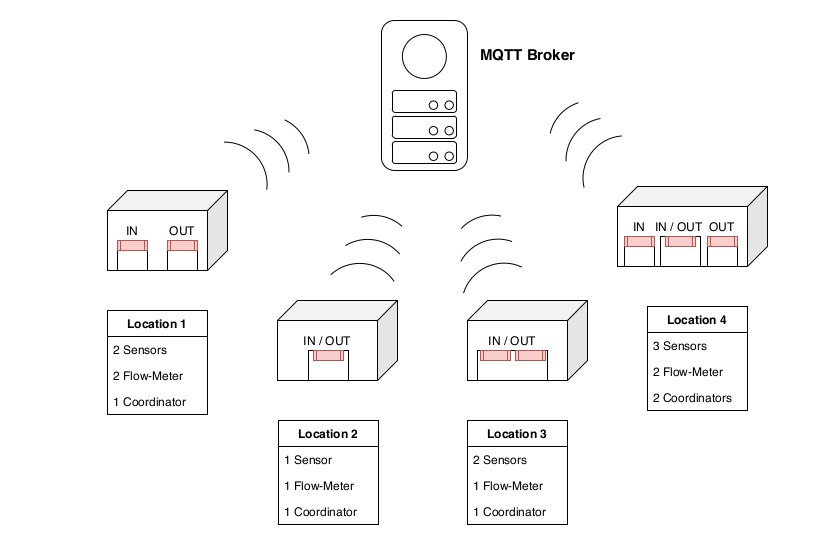}
        \label{fig:subfig6}}
    \hfill
    \subfloat[Data communication between MQTT broker and a coordinator]{
        \includegraphics[width=0.7\linewidth]{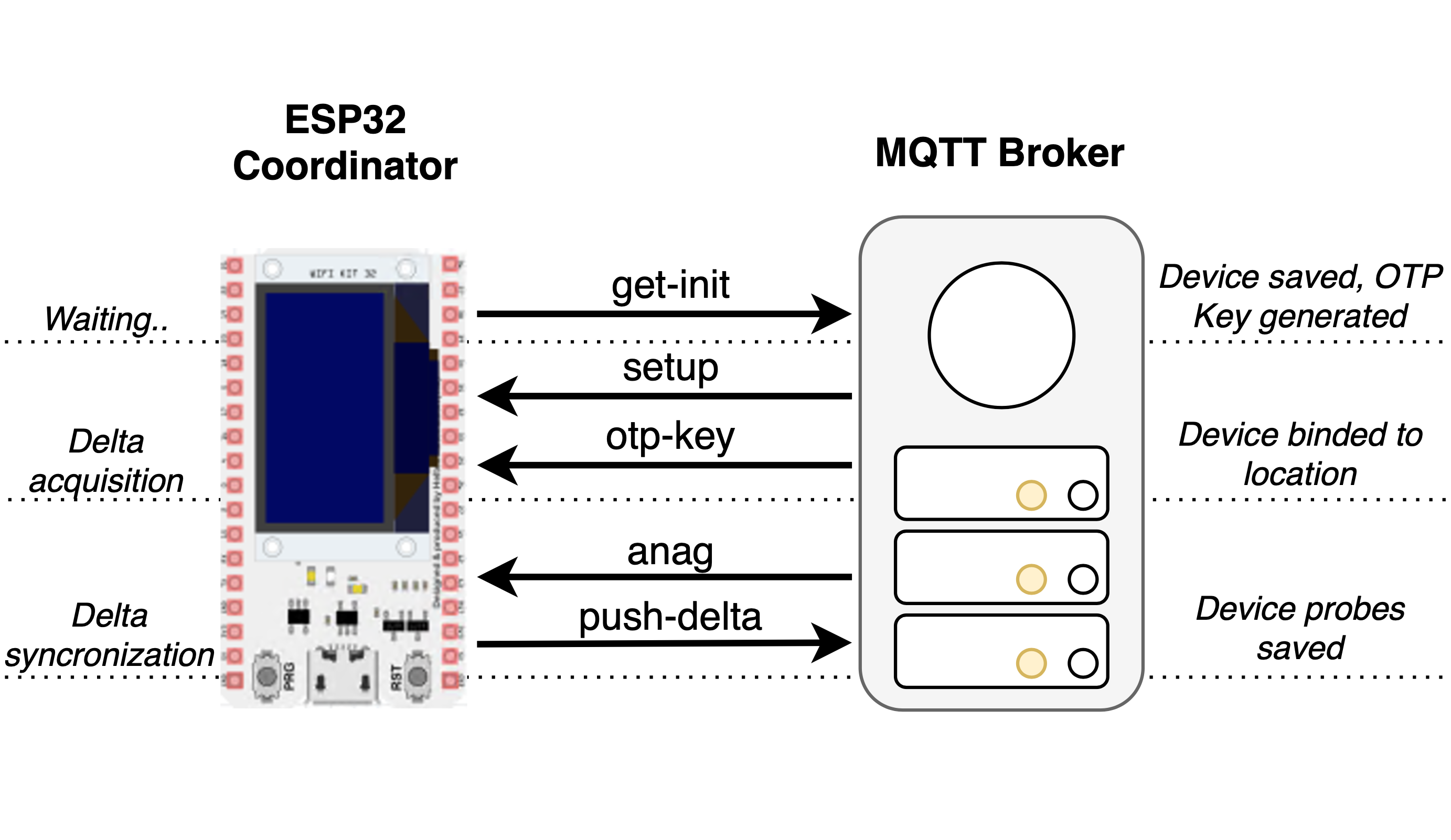}
        \label{fig:subfig7}}
    \caption{Coordination between microcontroller and broker}
    \label{figure:coordination}
\end{figure}

At this level, we can locate all of the equipment that deal with coordinating the data from the Flow-Meters. They are classified into two kinds:
\begin{itemize}
    \item within the infrastructure
    \item outside of the infrastructure
\end{itemize}
The former are \textit{microcontrollers that manage data synchronization} from many sorts of sensors.
The latter are servers (called \textbf{brokers}) that \textit{communicate with the former} through the MQTT protocol and ensure the consistency of infrastructure acquisitions.
Microcontrollers have the responsibility to manage this data in accordance with the standards of the linked devices and to act as an interface with the MQTT broker as shown in Figure \ref{fig:subfig6}.

Our infrastructure must be a closed system in order for the broker to be aware of the full flow.
This enables the broker to ensure that the incoming data is consistent and to store it in a database.
The broker and the coordinator use the MQTT protocol to interact safely and reliably, and they take particular actions based on the messages they receive as shown in Figure \ref{fig:subfig7}.

The connection initialization and initial configuration are performed during the first start, and the device is resumed once done.
During this phase, the device runs in WIFI\_STA\footnote{The possible working states are: WIFI\_AP (Access Point mode),WIFI\_STA (Client only), WIFI\_AP\_STA (dual mode), WIFI\_OFF (disabled)} modality and provides a setup web page via the \textit{Autoconnect library}, where the connection credentials may be entered.
On the server side, there is a validation and registration phase during the initial connection; from then on, at each successive start, the device requests the server for its personal data, publishing on the specific topic of the MQTT protocol.
The server responds with the device type, location information, and any setup constants, which are collected by three distinct callbacks.
If a single coordinator is responsible for managing a large number of sensors, a first coordination is necessary.
The influx of data from these sensors must be coordinated based on their characteristics; hence, the several Flow-Meters employ a common standard to transmit the specifications and results of their processing.
One of their parameters is the region the sensor can cover:
the objective is to broaden the range of action of certain types of sensors while keeping the system scalable, since signal processing and identification are delegated to a lower level, the Flow-Meter.
The coordinator then assigns a difference value to each of the sensors and updates it at each state change with the increase or reduction. Another thread is responsible for posting the updated value to the server.

At this level, coordination is based on an overview of the sites and
we expect that our sensors have mapped at least all of the entrance and exit areas.
The coordinators function as repeaters in the interaction with the Broker.
As a result, all coordinators within the location must be linked with it.
The broker must be aware of the whole flow, which occurs only if our site is a closed system; at that point, it has visibility of what comes in and what comes out.

Essentially, the protocol implemented provides us with channels to publish and receive messages among the various clients via callbacks. As a result, this paradigm does not provide intellectual ability for those that manage the messages, the broker, which is only limited to coordinating communication.
It was necessary to build an additional layer on top of an existing broker that would allow intercepting messages passing through the network and triggering server-side actions.
These activities include retrieving data, storing and maintaining information, and performing tasks based on the kind of message provided to the subjects, which are used as API endpoints.
At this stage, the server relays the message exchange, which does not just send the message but also applies predefined logic to it: an example might be the phase of adding a new device, which merely registers on the server. Following the authentication and authorization stages, the server verifies and tells the device about the place which it is allocated to, and then informs the other devices connected.

One of the challenges we faced throughout the project's execution was how to maintain the quantity of information arriving from the coordinators synced and the quality on a high level, because these devices, as previously said, are not so reliable when they perform this kind of tasks. To address this issue, it was therefore decided to adopt the MQTT protocol, which has a very high degree of reliability in handling the transmission of information across various devices and equipment.

Security deserves a separate mention, as there are two "by design" vulnerabilities of the MQTT protocol:
protection against unwanted subscription/publication and encryption of packets exchanged between client and broker.
Unwanted subscription protection can be achieved through a \textit{whitelist}. 
Each device is provided with a unique key and the subscription is made possible only if the key is valid. 
In the event that the key is compromised, a new one can be generated, as the device is capable of receiving updates via OTA (Over-The-Air).
By default, even to keep the protocol light, no encryption is applied to the packets.

There are several methods for securing message exchange and each of them should be chosen based on the device's power.
According to RFC 7228, these devices are classified into three types: 
our microcontroller belongs to the class 2 devices, capable of managing the overhead of the SSL/TLS protocol; additionally, because it is dual-core, it is possible to delegate the server update to a separate process that manages the queue, without slowing down the data acquisition process.

\subsection{Presentation Layer}
This level has the purpose of making the information stored to the user accessible through the devices located at the underlying levels, and to manage the access mode for single users, guaranteeing the separation of responsibilities for any of them.
So, main actions for this layer are:
\begin{itemize}
    \item make the information acquired by the devices located at the underlying levels available to the user
    \item manage access methods
    \item separate the responsibilities relating to the user profile
\end{itemize}
The first evident distinction is between the external user, called \textit{standard user}, who reads the information, and the internal user, called \textit{business}, who owns the information and determines how to display it.
The latter, unlike the former, must be registered in the system, in order to manage an activity.

\begin{figure}[h!]
    \centering
    \includegraphics[width=0.6\linewidth]{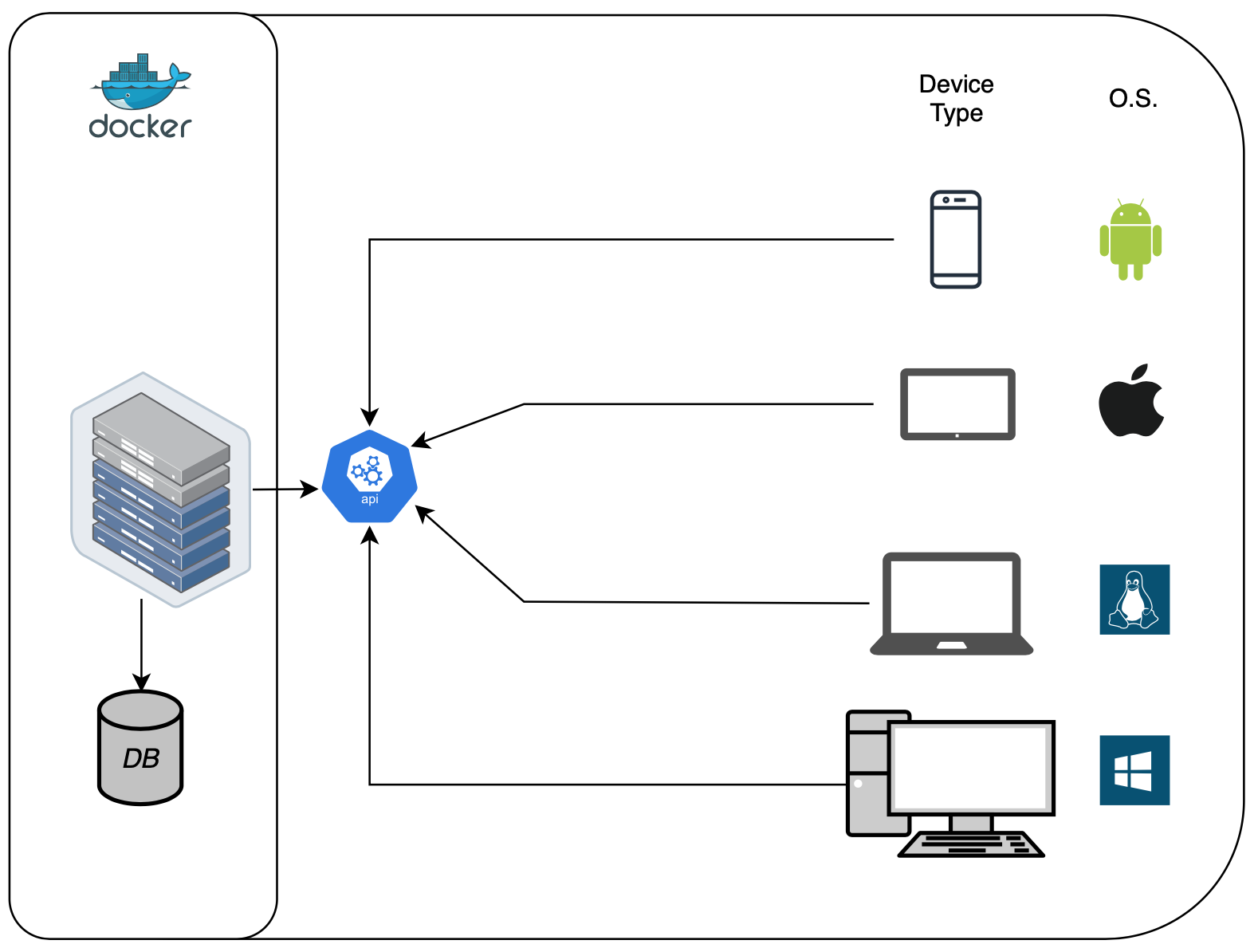}
    \caption{Presentation Layer and docker container}
    \label{fig:services}
\end{figure}

This level's structure has been extensively tested in various environments; there is a Docker container which virtualizes the data archived in the MySQL database in a secure environment, and the server, which runs on a virtual machine and externally exposing only the endpoints of the APIs that the application calls as shown in Figure \ref{fig:services}.

The presentation system is made up of two parts: the web server and the client.
Infrastructure services manage main database activities (CRUD) using APIs accessible to the application client.
These APIs are classified into three categories (authentication, task management, and device management), corresponding to a different set of entities in the database.

\begin{figure}[h!]
    \centering
    \includegraphics[width=0.6\linewidth]{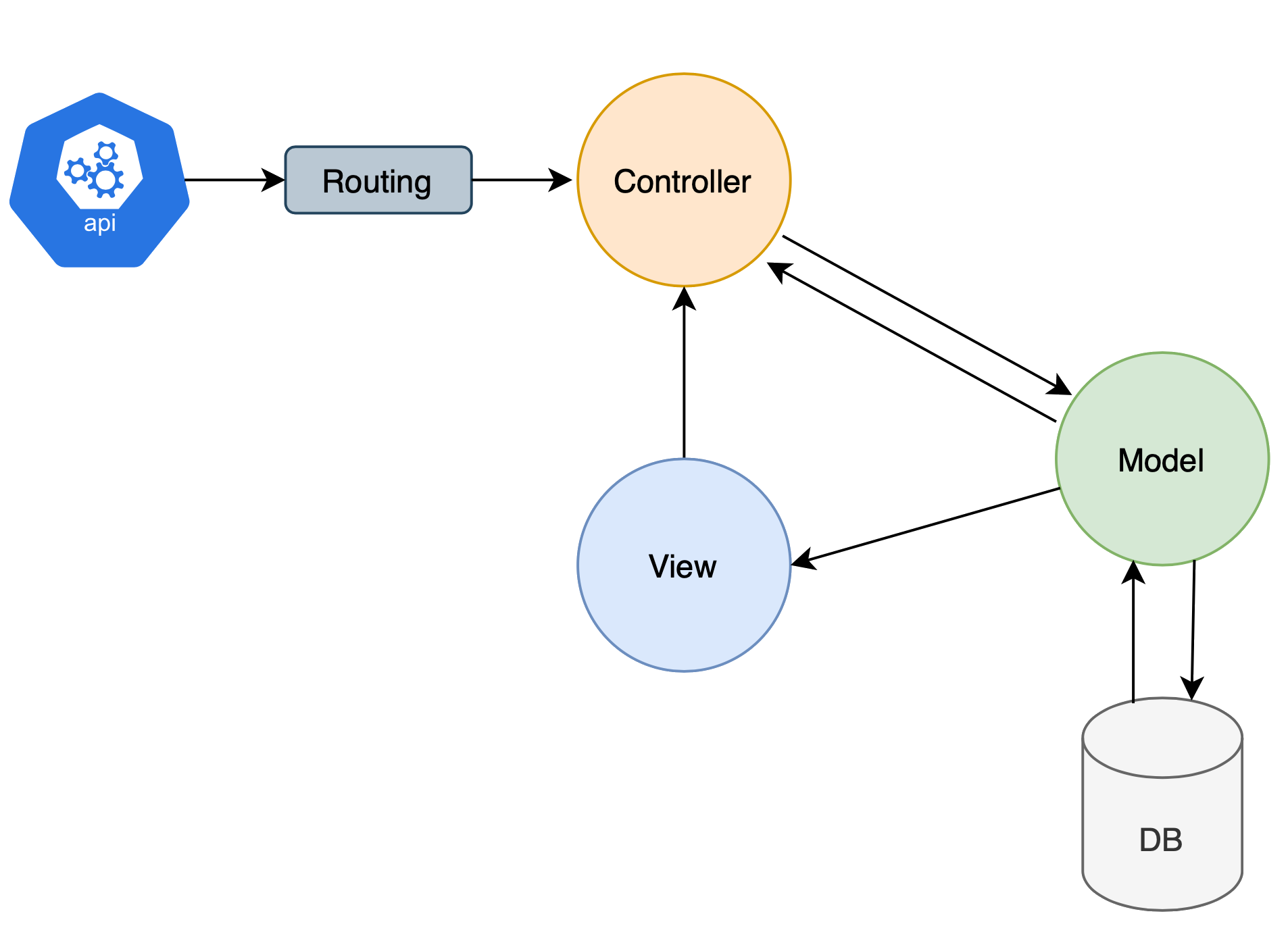}
    \caption{Scheme of MVC pattern}
    \label{fig:mvc}
\end{figure}

The development scheme we have chosen provides the \textit{Model View Controller} (MVC) pattern is an approach that involves categorizing the elements of an application into three distinct parts in order to make their modification and management autonomous, thereby increasing maintainability and portability.
These three categories are the \textit{Model}, the \textit{View}, and the \textit{Controller}, in that order as shown in Figure \ref{fig:mvc}.
\begin{itemize}
    \item \textit{Model}: contains the data access methods
    \item \textit{View}: takes care of viewing the data to the user and manages the interaction between the latter and the underlying infrastructure
    \item \textit{Controller}: receives user commands through the View and reacts by performing operations that may affect the Model and which generally lead to a change in the View state
\end{itemize}

Therefore, the \textit{Model level} is responsible for managing the data that the application must manipulate. The term Model implies its nature that is a representation of the object, which will subsequently be mapped to the database according to rules set by the programmer, abstracting the structure of the data within the database.

In the MVC framework, the \textit{Controller level} is the most crucial.
The Controller level contains all of the classes that allow you to validate user requests and process them in order to return a response.
It is able of intercepting user requests, choosing the components capable of processing them, mapping the result into an object, and returning it to the caller.

\begin{figure}[h!]
    \centering
    \includegraphics[width=0.8\linewidth]{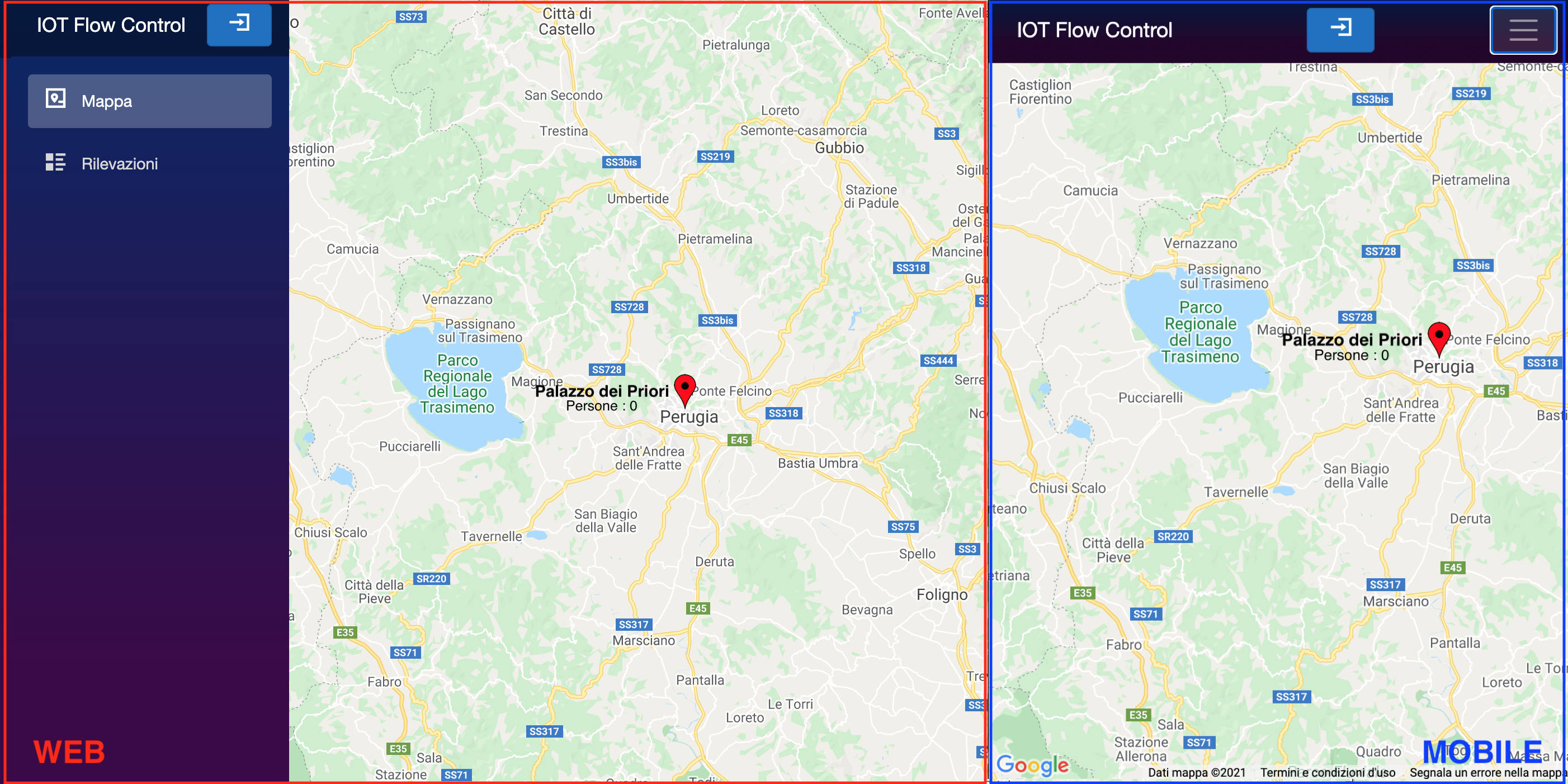}
    \caption{App responsive graphic interface}
    \label{fig:view1}
\end{figure}

To allow the user to interact with the system, we created a Progressive Web Application that can run within the browser and so adapt to any operating system or device, with a responsive UI.
The application's Home screen greets the user with a map that shows nearby areas of interest as well as associated information.
By picking the point of interest, you will have access to extra information that the management has decided to make available for the specific user as shown in Figure \ref{fig:view1}.

\begin{figure}[h!]
    \centering
    \includegraphics[width=0.8\linewidth]{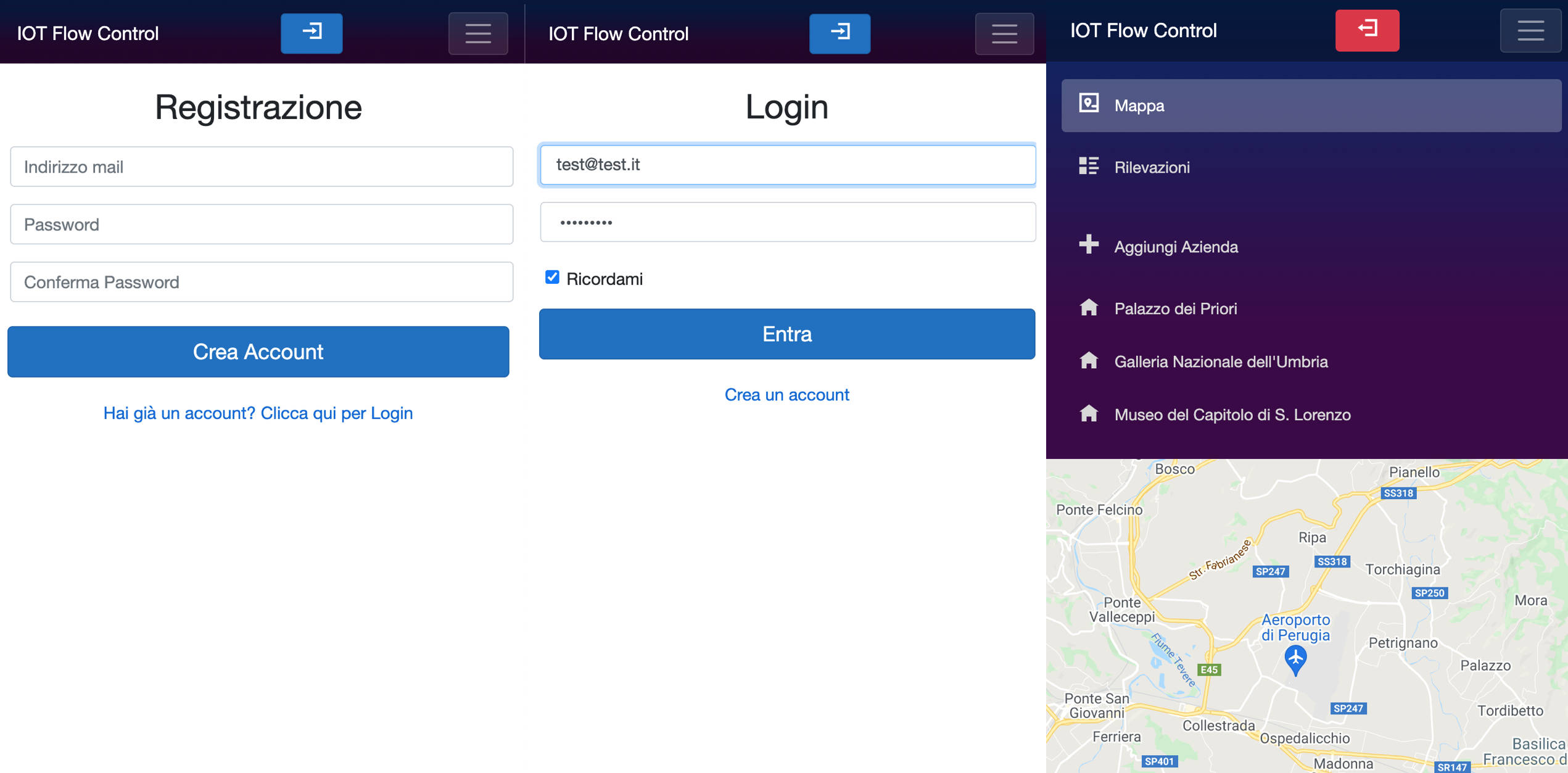}
    \caption{Registration, login and modification authorization phases}
    \label{fig:view2}
\end{figure}

The user must be registered and enabled in order to utilize the application's functions as a business role.
Every time the user takes an action, a request is sent to the server, and the request is successful if the formal check is accurate.
After an user has been verified and approved, he/she may manage their corporate profiles as shown in Figure \ref{fig:view2}.
To publish updates on their activity, the user has to enter them and associate the devices with them too.
The localization information about the site is retrieved through a \textit{Geocoding} service.
When the insertion is finished, the server changes the tables and notifies the other clients via \textit{RPC} and so the device is completely associated with the support of an \textit{OTP Key}.

The application has also been enriched with an additional module that allows the visitor to obtain information on artifacts where appropriate (museums, exhibitions, etc.).
With the help of virtual reality techniques, the user can take a guided tour of the facility and view the main works of art.
Furthermore, a series of QR codes have been created inside the structure that allow the user to view the artifact through augmented reality technologies.

\section{Discussion of Results}
We proceeded with the verification of the overall correctness of the system after completing all of the steps mentioned in the previous chapters. All testing were carried out on a precised room within the University area with a single entrance and exit door, with a setup consisting of a single Grid-EYE sensor, a Flow-Meter, and a Coordinator. We tested the system in its entirety after performing all of the tests required to ensure the system's correct operation and, later, those required to refine the functionality of the individual components. 
The following operations were then performed:
\begin{enumerate}
     \item Positioning of the Grid-EYE sensor on the door jamb, at a height of about two meters
     \item Connection of the sensor to the acquisition microcontroller, which in turn is connected to the coordinator
     \item Connecting the microcontrollers to the network
     \item Configuration of Coordination Layer (Broker MQTT) and Presentation Layer (database and web server) on two separate machines
     \item Installing the application on an Android device
     \item Registration via email and password
     \item Authentication and authorization within the system
     \item Insertion of a fictitious activity so that it can appear on the map and be selected from the context menu
     \item Association of the coordinator via the OTP Key. The correct name for a business role appeared on the microcontroller display and the device appeared on the proper page.
\end{enumerate}

At this point, the activity and the device had been properly setup, and the fifteen-day testing period could begin.
We turned on the detector at the start of the first day, when all staff (professors, students, and so on) members were not yet in the room.
We found that every time a person went in, the counter was incremented and when he/she went out, the number was decremented.
At the conclusion of the test period, late in the last day's evening, the counter was correctly returned to zero.

\section{Conclusions and Future Works}
The testing revealed that our system has a strong basic detection capacity, as we discovered a result of zero for eight days, a value of one for four days, and a value of two for the remaining three days on 42 average daily passes. 
The detections were monitored using the program, which checked that the inputs and outputs were appropriately synchronized whenever a state change happened. 
It was also able to examine the logs, which revealed that the mistakes were caused by erroneous assessments by the acquisition tools. 
The different test phases have demonstrated how, although being in an experimental phase, our technology seems to be adequate for creating a reliable infrastructure.
It has been shown that the Internet of Things provides nearly limitless tools and potential, both in terms of opportunities and future extensions, ranging from the development of a simple information process to the implementation of a real portal capable of allowing those who do business to communicate with interested customers. 
An exciting future use might be to make available, in addition to the actual survey summary, booking systems capable of sending the flow of people, which, when combined with the information status signals, allows customers to have direct contact with who provides the service in real time.

\addcontentsline{toc}{section}{Acronyms}
\section*{Acronyms}
The following acronyms are used in this manuscript:

\noindent
\setlength{\tabcolsep}{15pt}
\renewcommand{\arraystretch}{1.2}
\begin{tabular}{@{}ll}
OTP & One Time Password \\
RPC & Remote Procedure Call \\
CRUD & Create, Read, Update, Delete \\
MVC & Model View Controller \\
UI & User Interface \\
CCTV & closed-circuit TV \\
PIR & Passive InfraRed \\
IOT & Internet of Things \\
RAM & Read Only Memory \\
CPU & Central Processing Unit \\
OTA & Over The Air \\
API & Application Program Interface \\
MQTT & MQ Telemetry Transport or Message Queue Telemetry Transport \\
\end{tabular}
\vskip 1cm

\printbibliography
s
\end{document}